# Effect of zirconium doping on mechanical properties of $W_{1-x}Zr_xB_2$ on the base of *ab initio* calculations and magnetron sputtered films.

Marcin Maździarz[1*], Rafał Psiuk[1], Agnieszka Krawczyńska[2], Małgorzata Lewandowska[2] and Tomasz Mościcki[1]

[1]Institute of Fundamental Technological Research Polish Academy of Sciences, Pawińskiego 5B, Warsaw, 02-106, Poland.
[2]Warsaw University of Technology, Faculty of Materials Science and Engineering, Wołoska 141, Warsaw, 02-507, Poland.

*Corresponding author(s). E-mail(s): mmazdz@ippt.pan.pl;
Contributing authors: rpsiuk@ippt.pan.pl;
agnieszka.krawczynska@pw.edu.pl;
malgorzata.lewandowska@pw.edu.pl; tmosc@ippt.pan.pl;

**Abstract**

Potentially superhard $W_{1-x}Zr_xB_2$ polymorph hP6-P6₃/mmc-WB₂ with zirconium doping in the range of x= 0-25% was thoroughly analyzed within the framework of first-principles density functional theory from the structural and mechanical point of view. The obtained results were subsequently compared with properties of material deposited by magnetron sputtering method. All predicted structures are mechanically and thermodynamically stable. Due to theoretical calculations zirconium doping reduces hardness and fracture toughness $K_{IC}$ of $WB_2$. Deposited films are characterized by greater hardness $H_v$ but lower fracture toughness $K_{IC}$. The results of experiments show that not only solid solution hardening is responsible for strengthening of predicted new material but also change of microstructure, *Hall–Petch* effect and boron vacancies.

# 1 Introduction

The need to discover new materials is a scientific and industrial topic covering all areas of application. Recent studies on superhard materials have shown that computational-based understanding and modeling serves as a reliable trend indicator and can be used to experimental design of new materials or their special properties [1–3]. Transition metal borides are a fairly of new and extremely promising class of materials that can be used in a wide variety of applications, from superconductivity to wear and corrosion resistance [4–7]. Unlike nitrides or carbides, the knowledge on these materials is not very large and requires further research, especially about compounds like ternary borides with improved ductility and increased crack resistance correlated with great hardness. Although there are experimental and theoretical studies on binary borides [8], a tungsten diboride doped with transition metals is rather unexplored.

Particularly noteworthy are ternary metals doped tungsten borides. Recent experimental studies show that due to the addition of a lighter element than tungsten, the specific density of the obtained material can be reduced with the maintaining of the very high hardness and thermal stability. Weight reduction is crucial with commercially used tungsten carbides, for example. In addition, during synthesis of this ternary borides hexagonal borides like $TiB_2$ and $ZrB_2$ (P6/mmm space group) can be created [9]. They have very high melting point and at the same time are characterized by high thermal and electrical conductivity [8, 10]. Therefore, the introduction of tantalum or zirconium into the $WB_2$ crystal lattice can give the opportunity to create a new group of hard and refractory materials. Hitherto, the $W-Ta-B_2$ system has possessed greater interest of researchers [3, 11, 12]. The theoretical and experimental studies have been mainly concerned on the obtaining of this material in the form of coatings. It has been shown that for different tantalum and tungsten contents magnetron deposited thin films of $Ta_xW_{1-x}B_{2-z}$ crystallize mainly in the alpha type structure. Experimental studies have shown that the obtained layers are characterized by great hardness of around 45 GPa and fracture toughness $K_{IC}$ values 3.0 MPa$\sqrt{m}$ what make this compound much better than common $TiN$, $Ti-Si-N$ and $(Ti,Al)N$ [11]. Also, the good thermal stability and high oxidation resistance at temperatures up to 700°C [12] makes them a promising candidate for industrial applications. Additional function of theoretical calculations is possibility of explanation of material properties origin. Using DFT *ab initio* methods for example, it has been shown that vacancies are responsible for metastable $\alpha$-phase stabilization in ternary system [3]. It should be emphasized that earlier theoretical calculations questioned the possibility of the existence under ambient conditions of the metastable $\alpha-WB_2$ phase (space group 191 (P6/mmm)) with a hexagonal, similar to $AlB_2$ structure and came to the conclusion that it should be a high-pressure phase which can be stable above 65 GPa [13, 14]. The latest studies have been shown that tungsten diboride alloyed with zirconium also possess great potential. Deposited by Radio Frequency (RF) magnetron sputtering $W_{0.8}Zr_{0.2}B_{1.9}$ coatings are super-hard with a hardness reaching 43.9± 3.3 GPa and possess the fracture toughness $K_{IC}$ values of 1.77 MPa$\sqrt{m}$ [15], what is comparable with much softer *TiN* [16]. Deposited on high speed steel coatings are smooth $R_a$= 0.024 $\mu$m and also possess good adhesion. The estimated in this case amplitude of the tensile wave that caused coating delamination was about 320 MPa [17]. Psiuk et al. [18] proposed combined magnetron sputtering and pulsed laser deposition technique for doping $WB_2$ films by zirconium. By changing the laser fluence at the surface of zirconium diboride target it was possible to control the ablation rate and hence the dopant content in deposited film. In consequence, the microstructure and mechanical properties of RF magnetron sputtered $WB_2$ coatings were changed. Film obtained with the fluence 1.06 J/cm$^2$ (~2% *Zr* at.) shows ductile–brittle behavior and are superhard $H_v$ = 40±4 GPa, incompressible $R_s$ = 12±1 GPa, and possess relatively low Young's modulus $E$ = 330±32 GPa and high elastic recovery $W_e$ = 0.9 [18]. In the case of plasma sintered (fast-SPS) (*W*, *Zr*)$B_2$ compacts the obtained Vickers hardness values measured at 1 N was 24.8± 2.0 for 24 at.% zirconium in $WB_{2.5}$ $W_{0.76}Zr_{0.24}B_{2.5}$ specimens showed electrical conductivity up to 3.961·10$^6$ S/m, which is higher than stainless steels and similar to *WC–Co* cemented carbides [19]. Additionally, the SPSed specimens were characterized by a high specific wear rate of 3·10$^{-5}$ mm$^3$/Nm. The friction and wear test results revealed the formation of a boron-based film which seems to play the role of solid lubricant. In other experimental and theoretical studies on $Zr_xW_{1-x}B_2$ diborides zirconium content of *x*>0.24 was analyzed. Gu et al. [20] report on a systematic first-principles study of a large series of group-IVB, VB and VIB dual-TM diborides in hexagonal structure to explore the brittle-ductile relation. For $W_{0.75}Zr_{0.25}B_2$ compound the starting structure was hexagonal $ZrB_2$ in P6/mmm symmetry, which is the hardest, stable polymorph of zirconium diboride obtained from theoretical calculations [21]. The replacement of 6 from 8 zirconium atoms with tungsten resulted in a significant drop in hardness ($H_v$ = 17.44 GPa). At the same time, this compound is characterized by a relatively high Poison's ratio $v$ = 0.268 and a low Pugh's ratio G/B = 0.548 (ratio of shear and bulk modulus), which qualifies it as a ductile material. The presented so far experimental studies have been aimed at improving the mechanical properties of the obtained material and the maintaining of very good thermal properties of $ZrB_2$ [22–26]. Ordan'yan et al. [26] presented that $W_2B_5$ and $ZrB_2$ form a eutectic at 20 mol% $ZrB_2$ with a melting point of 2180°C. Microhardness and diffraction tests showed the hardness and lattice parameters practically do not differ in the whole solid solution $W_2B_5$–$ZrB_2$ system. Experimental investigations on lower concentration of zirconium *x*<0.5 was also carried out for phases with higher content of boron, ie. tungsten tetraboride. As it has been presented in [9], the lower content of zirconium makes the synthesized $W_{1-x}Zr_xB_4$ materials harder than the non-doped tungsten tetraborides and are more resistant to oxidation. In the case of $W_{0.92}Zr_{0.08}B_4$ the hardness 34.7±0.65 GPa under an applied load of 4.9 N was measured. It is the highest value obtained for any superhard metal ceramics at this relatively high loading [9]. In the case of low ($x<0.25$) zirconium content, no theoretical studies have been performed so far.

The aim of this work is using of density functional theory (DFT) calculations to determine the effect of doping tungsten diborides with transition metal such as zirconium to characterize its structure and mechanical properties. The comparison with experiment will try to explain possible mechanisms of hardening of magnetron sputtered W−Zr−B coatings which can be used with success in nowadays engineering.

# 2 Methodology

## 2.1 Computational Methodology

### 2.1.1 *Ab Initio* Calculations

First-principles calculations based on density functional theory (DFT) [27, 28] within the pseudopotential plane-wave approximation (PP-PW) implemented in ABINIT [29, 30] software were carried out in this study. Projector augmented-wave formulation (PAW) pseudopotentials [31] were employed to represent the interactions of ionic core and non-valence electrons.

The effect of choice of an exchange-correlation (XC) functional on calculated lattice constants in $WB_x$ structures was analyzed in [32]. Analysis of the experimental data [14] suggests that it is reasonable to use local density approximation (LDA) [33, 34] as a XC functional. Projector augmented wave method (PAW) pseudopotentials used for LDA XC functionals were obtained from PseudoDojo project [35]. The following valence electron configurations were used: the $5s^2 5p^6 5d^4 6s^2$ for $W$, the $2s^2 2p^1$ for $B$ and the $4d^2 5s^2$ for $Zr$, respectively.

The calculation accuracy settings correspond to those in the work [21].

### 2.1.2 Generation and Optimization of Structures

Tungsten borides crystallize in various phases but $WB_2$ hP6-P6$_3$/mmc (194) seems to be the hardest one [32, 36] and how the addition of zirconium affects the mechanical properties of this phase, see Fig.1, was investigated. In order to do this the following supercells of $WB_2$: $2\times 1\times 1$ (12 atoms), $3\times 1\times 1$ (18 atoms), $2\times 2\times 1$ (24 atoms), $3\times 2\times 1$ (36 atoms), $2\times 2\times 2$ (48 atoms), $3\times 3\times 1$ (54 atoms), $3\times 2\times 2$ (72 atoms) and $4\times 4\times 1$ (96 atoms) were generated and one arbitrary tungsten atom was replaced with a zirconium atom. An example of the generated supercell is shown in Fig.2. Which atom is chosen does not matter, they are equivalent. However, the doped structure has a different symmetry than the symmetry of the original $WB_2$ one, see Tab.1. The generated structures were then fully optimized, cell geometry and atomic coordinates, just as in [21, 37].

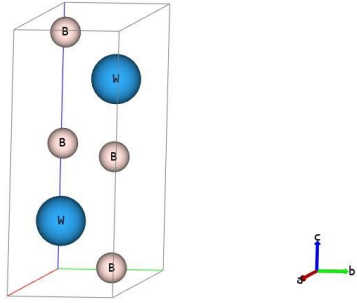

**Fig. 1** Basic cell of $WB_2$: hP6-P6$_3$/mmc (194)

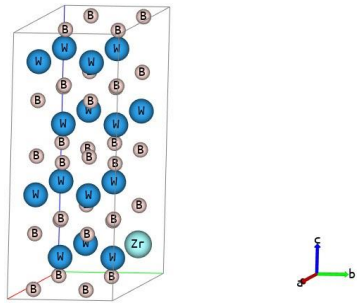

**Fig. 2** 2×2×2 supercell with one W atom replaced by Zr atom $W_{0.94}Zr_{0.06}B_2$ ($W_{15/16}Zr_{1/16}B_2$)

### 2.1.3 Formation Enthalpy and Cohesive Energy

The formation enthalpy and cohesive energy of $W_{1-x}Zr_xB_2$ structures were determined as follows [38, 39]:

$$\Delta_f H(W_{1-x}Zr_xB_2) = E_{coh}(W_{1-x}Zr_xB_2) - (1-x)E_{coh}(W) - xE_{coh}(Zr) - 2E_{coh}(B), \quad (1)$$

$$E_{coh}(W_{1-x}Zr_xB_2) = E_{total}(W_{1-x}Zr_xB_2) - (1-x)E_{iso}(W) - xE_{iso}(Zr) - 2E_{iso}(B), \quad (2)$$

where $\Delta_f H(W_{1-x}Zr_xB_2)$ is the formation enthalpy of the $W_{1-x}Zr_xB_2$; $E_{coh}(W_{1-x}Zr_xB_2)$ is the cohesive energy of the $W_{1-x}Zr_xB_2$; $E_{coh}(W)$ is the cohesive energy of W; $E_{coh}(Zr)$ is the cohesive energy of Zr; $E_{coh}(B)$ is the cohesive energy of B; $E_{tot}(W_{1-x}Zr_xB_2)$ is the total energy of the $W_{1-x}Zr_xB_2$; $E_{iso}(W)$ is the total energy of a W atom, $E_{iso}(Zr)$ is the total energy of a Zr atom and $E_{iso}(B)$ is the total energy of a B atom.

To calculate the cohesive energy as reference structures were chosen: for tungsten (cF4-Fm-3m (225)), for zirconium (hP2-P6/mmc (194)) and for boron (hR12-R-3m (166)).

### 2.1.4 Mechanical Properties Calculations

The theoretical ground state elastic constants $C_{ij}$ of all analyzed structures were calculated using the metric tensor formulation of strain in density functional perturbation theory (DFPT) [40]. Isotropised bulk modulus $B$, shear modulus $G$, Young's modulus $E$ and Poisson's ratio $\nu$ were estimated using a *Voigt–Reuss–Hill* average [41, 42].

In order to verify the mechanical stability of all the structures, positive definiteness of the stiffness tensor was examined [43] by calculating Kelvin moduli, i.e. eigenvalues of stiffness tensor written in *second-rank tensor* notation [44].

Hardness $H_v$ and fracture toughness $K_{IC}$ of all $W_{1-x}Zr_xB_2$ samples analyzed were estimated with the use of semi-empirical formulas developed in [45].

The brittle or ductile behavior of the material was examined on the basis of the Pugh's ratio $B/G$ [46]. Flexibility of hard nanocomposite coatings was estimated by $H_v/E*$ ratio, where $E* = E/(1 - \nu^2)$ [47].

## 2.2 Experimental methods

### 2.2.1 Process of magnetron sputtering

The ternary sputtering targets with a diameter of 25.4 mm were produced by Spark Plasma Sputtering (SPS) process from boron (purity: 95%, average particle size APS: 1 $\mu$m, Sigma Aldrich), tungsten (purity: 99.9%, APS: 25 $\mu$m, Sigma Aldrich), and zirconium (purity: 99.8%, APS: 250-350 $\mu$m, Sigma Aldrich). The composition of used for deposition targets was $WB_{2.5}$, $W_{0.92}Zr_{0.08}B_{2.5}$, $W_{0.84}Zr_{0.16}B_{2.5}$, $W_{0.76}Zr_{0.24}B_{2.5}$. Detailed information on SPS targets is presented in Ref.[19].

The target was mounted in the water-cooled 1-ich magnetron sputtering cathode (Kurt J. Lesker). The deposition process occurred in a vacuum chamber initially pumped to $2 \cdot 10^{-5}$ mbar and then filled with an argon to working pressure $9 \cdot 10^{-3}$ mbar. The gas flow of argon was 19 mL/min. Prior to each deposition, the target was sputtered for 5 min in order to ensure its clean surface and stable sputtering conditions. During all experiments, power supplied to the magnetron cathode was maintained at 50 W. Films were deposited for 180 min on $Si$ (100) (Institute of Electronic Materials Technology, Poland) and nitrided QRO90 steel substrates heated up to 450°$C$ and positioned 40 mm in front of the target. The deposited coatings were about 2.8 $\mu$m thick (Fig. 6a).

### 2.2.2 Characterization

The surface, cross section as well as chemical composition were investigated using scanning electron microscope (SEM) Hitachi Su8000. The chemical composition were investigated using Scanning Electron Microscope – SEM (JEOL JSM-6010Plus) equipped with Energy Dispersive X-Ray Spectroscope (EDS). Microstructural studies were done on the cross sections of deposited films cut perpendicularly to the surface. Electron transparent samples were prepared by focus ion beam (FIB) to a thickness of ≈300 nm. Acceleration voltage of the beam was set to 40 kV, while the beam current was controlled with the size of aperture depending on the stage of preparation. Then, lamellas were gently thinned by a low energy $Ar+$ ion beam system (between 0.6 and 1 keV) to a final thickness of ≈100 nm to limit additional defects presence in the microstructure introduced by the FIB beam. General observations were done with the use of Hitachi HD2700 scanning transmission electron microscope (STEM) operated at 200 kV. The STEM images were taken in bright-field (BF) and selective area electron diffraction (SAED) mode. In SAED mode the patterns were acquired by inserting a 250 nm of aperture. During the chemical composition of deposited coatings measurements accelerating voltage 5 kV was used. Moreover, the system was calibrated with the use of commercially available target $W_2B_5$ (purity 99.9%, Huizhou Tian Yi Rare material Co. Ltd). The authors are aware of the uncertainties in boron measurement with EDS, which are related to the proximity of the boron and carbon peaks as well as carbon contamination. The phase composition and crystal structure of deposited layers were characterized by X-Ray Diffractometer (Bruker D8 Discover, $\lambda$=1.5418Å). Measurements were taken in 2$\Theta$ scan mode, with source fixed at 8° position. In this configuration, it was possible to avoid signal from the substrate while maintaining high intensity of the signal originating primarily from the coating.

### 2.2.3 Mechanical properties

Vickers hardness measurements were performed using Wilson VH1102 microhardness tester (Buehler, Lake Bluff, IL, USA). Loads of 10 g were used to measure hardness of deposited coatings, 10 indentations on each sample were performed. For lower loads, a laser confocal microscope VK-X100 (Keyence, Osaka, Japan) was used to measurement of the indents. Laugier model formula [48] was used to analyze changes in fracture toughness ($K_{IC}$). For each coating, based on a group of 10 indentations, the mean value of $l$ and a crack dimensions, were determined similarly to [49]. Equation 3 was used to evaluate the fracture toughness of deposited coatings.

$$K_{IC} = x_v \left(\frac{a}{l}\right)^{\frac{1}{2}} \left(\frac{E}{H_v}\right)^{\frac{2}{3}} \frac{P}{c^{\frac{3}{2}}}, \quad (3)$$

where $K_{IC}$ is fracture toughness (MPa$\sqrt{m}$); $x_v$-indenter geometry factor (for Laugier Equation $x_v$=0.016); $E$-Young modulus of coating (GPa) (values were taken from [50]); $H_v$-hardness of coating (GPa); $P$ –the indentation load (mN); $a$-the length from the center of the indent to the corner of the indent (m); $l$-the length of the cracks; $c = l + a$.

# 3 Results and discussion

## 3.1 DFT calculations

The resulting structures obtained from the optimization are summarized in Tab.1 whereas crystallographic data as crystallographic information files (CIFs) and their figures are attached in the Appendix. Four samples of $W_{1-x}Zr_xB_2$ have a space group (Amm2 (38), orthorhombic crystal system), three (Pm (6), monoclinic crystal system), one (Pmm2 (25), orthorhombic crystal system) and initial $WB_2$ (P6$_3$/mmc (194), hexagonal crystal system). As expected, the decrease in the proportion of $Zr$ in the structures is followed by a decrease of average atomic volume, simply because the $Zr$ atom is "bigger" than $W$. The opposite trend is observed for formation enthalpy $\Delta_f H$ and cohesive energy $E_c$, see Fig.3. Please note that there are negative values in Tab.1 and Fig.3. This suggests that doped structures are less thermodynamically stable than pure ones, but still stable due to the negative value of $\Delta_f H$. Similar behavior was also observed for doping with $Ti$, $Al$, $V$, see [51, 52].

**Table 1** Chemical formula; Space group; Pearson symbol; proportion of $Zr$ dopant: $Zr/(W + Zr)$; Volume per atom (Å/Atom); formation enthalpy $\Delta_f H$ (eV/Atom); cohesive energy $E_c$ (eV/Atom)

| Sample | Space group | Pearson symbol | $\frac{Zr}{W+Zr}$ | Vol/atom | $-\Delta_f H$ | $-E_c$ |
|---|---|---|---|---|---|---|
| $WB_2$ | P6$_3$/mmc (194) | hP6 | 0/1 | 9.298 | 0.495 | 8.880 |
| $W_{0.75}Zr_{0.25}B_2$ | Pmm2 (25) | oP12 | 1/4 | 9.775 | 0.395 | 8.300 |
| $W_{0.835}Zr_{0.165}B_2$ | Pm (6) | mP18 | 1/6 | 9.615 | 0.423 | 8.488 |
| $W_{0.875}Zr_{0.125}B_2$ | Amm2 (38) | oC24 | 1/8 | 9.533 | 0.438 | 8.583 |
| $W_{0.915}Zr_{0.085}B_2$ | Pm (6) | mP36 | 1/12 | 9.455 | 0.455 | 8.681 |
| $W_{0.937}Zr_{0.063}B_2$ | Amm2 (38) | oC48 | 1/16 | 9.417 | 0.466 | 8.731 |
| $W_{0.944}Zr_{0.056}B_2$ | Amm2 (38) | oC54 | 1/18 | 9.403 | 0.468 | 8.747 |
| $W_{0.958}Zr_{0.042}B_2$ | Pm (6) | mP72 | 1/24 | 9.381 | 0.475 | 8.780 |
| $W_{0.969}Zr_{0.031}B_2$ | Amm2 (38) | oC96 | 1/32 | 9.360 | 0.479 | 8.805 |
| W | Fm-3m (225) | cF4 | | 15.78 | | 12.64 |
| Zr | P6$_3$/mmc (194) | hP2 | | 22.30 | | 6.87 |
| B | R-3m (166) | hR12 | | 20.95 | | 6.26 |

The calculated Kelvin moduli, i.e. eigenvalues of stiffness tensor, for all analyzed structures are given in Tab.2. It can be seen that all and for each sample are positive. This means mechanical stability of all the structures.

The symmetry of a crystal determines the symmetry of its physical properties, here we are interested in the symmetry of the stiffness tensor and the number of distinct elastic constants [53]. For hexagonal crystal system we have 5, for the orthorhombic 9 and for the monoclinic 13 distinct elastic constants, respectively. For clarity of presentation we have included the full stiffness tensors for each structure in the Appendix. The derived quantities from the elasticity constants are listed in Tab.3. An increase in the proportion of $Zr$ in $W_{1-x}Zr_xB_2$ reduces the value of Bulk modulus $B$, shear modulus $G$,

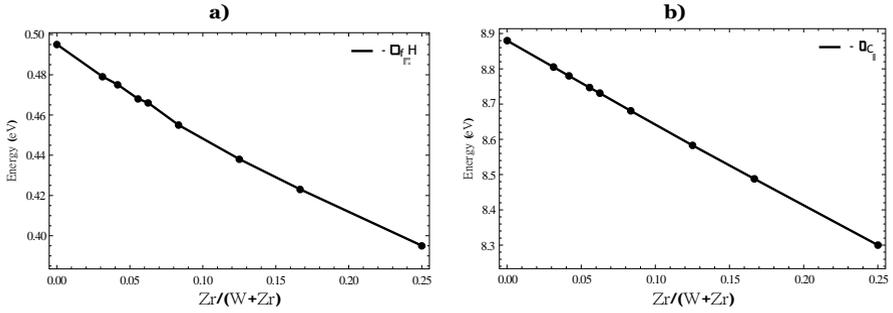

**Fig. 3** $W_{1-x}Zr_xB_2$: a) formation enthalpy $\Delta_f H$ (eV/Atom), b) cohesive energy $E_c$ (eV/Atom)

**Table 2** Chemical formula; Kelvin moduli $K_i$ (GPa)

| Sample | $K_1$ | $K_2$ | $K_3$ | $K_4$ | $K_5$ | $K_6$ |
|---|---|---|---|---|---|---|
| $WB_2$ | 1065.21 | 705.223 | 552.611 | 552.611 | 403.475 | 403.475 |
| $W_{0.75}Zr_{0.25}B_2$ | 873.269 | 556.857 | 401.992 | 398.155 | 287.685 | 286.955 |
| $W_{0.835}Zr_{0.165}B_2$ | 946.392 | 602.807 | 449.931 | 435.294 | 324.303 | 304.977 |
| $W_{0.875}Zr_{0.125}B_2$ | 963.91 | 627.05 | 463.626 | 459.192 | 336.276 | 333.599 |
| $W_{0.915}Zr_{0.085}B_2$ | 972.758 | 600.622 | 504.494 | 456.625 | 319.321 | 310.069 |
| $W_{0.937}Zr_{0.063}B_2$ | 996.274 | 651.301 | 487.159 | 485.881 | 351.937 | 351.842 |
| $W_{0.944}Zr_{0.056}B_2$ | 1025.74 | 652.494 | 504.358 | 503.768 | 358.926 | 350.14 |
| $W_{0.958}Zr_{0.042}B_2$ | 1008.52 | 642.412 | 505.422 | 496.438 | 352.005 | 339.054 |
| $W_{0.969}Zr_{0.031}B_2$ | 1023.54 | 645.426 | 509.716 | 508.957 | 350.915 | 346.937 |

Young's modulus $E$, hardness $H_v$ and fracture toughness $K_{IC}$. The Poisson's ratio $v$ is nearly constant at around 0.2. The Pugh's ratio $B/G$ is a relationship associated with brittle or ductile behavior of materials. Higher $B/G$ ratio corresponds to higher ductility [46]. In our case, the doping of $Zr$ increases ductility and at the same time reduces the $H_v/E^*$ ratio, i.e. some measure of flexibility of hard nanocomposite coatings.

**Table 3** Chemical formula; Bulk modulus $B$ (GPa); shear modulus $G$ (GPa); Young's modulus $E$ (GPa); Poisson's ratio $v$; $B/G$ Pugh's ratio; hardness $H_v$ (GPa); hardness to modified Young's modulus ratio $H_v/E^*$; fracture toughness $K_{IC}$ (MPa $\sqrt{m}$)

| Sample | $B$ | $G$ | $E$ | $v$ | $B/G$ | $H_v$ | $H_v/E^*$ | $K_{IC}$ |
|---|---|---|---|---|---|---|---|---|
| $WB_2$ | 336.64 | 258.99 | 618.40 | 0.19 | 1.30 | 34.27 | 0.053 | 5.46 |
| $W_{0.75}Zr_{0.25}B_2$ | 273.45 | 190.19 | 463.19 | 0.22 | 1.44 | 22.20 | 0.046 | 3.78 |
| $W_{0.835}Zr_{0.165}B_2$ | 295.13 | 208.71 | 506.68 | 0.21 | 1.41 | 24.71 | 0.047 | 4.27 |
| $W_{0.875}Zr_{0.125}B_2$ | 305.66 | 218.50 | 529.36 | 0.21 | 1.40 | 26.16 | 0.047 | 4.52 |
| $W_{0.915}Zr_{0.085}B_2$ | 296.39 | 216.42 | 522.16 | 0.21 | 1.37 | 26.56 | 0.049 | 4.37 |
| $W_{0.937}Zr_{0.063}B_2$ | 313.63 | 229.70 | 553.88 | 0.21 | 1.37 | 28.30 | 0.049 | 4.76 |
| $W_{0.944}Zr_{0.056}B_2$ | 319.65 | 234.22 | 564.73 | 0.21 | 1.36 | 28.87 | 0.049 | 4.90 |
| $W_{0.958}Zr_{0.042}B_2$ | 311.88 | 230.88 | 555.55 | 0.20 | 1.35 | 28.85 | 0.050 | 4.77 |
| $W_{0.969}Zr_{0.031}B_2$ | 315.23 | 233.78 | 562.32 | 0.20 | 1.35 | 29.29 | 0.050 | 4.83 |

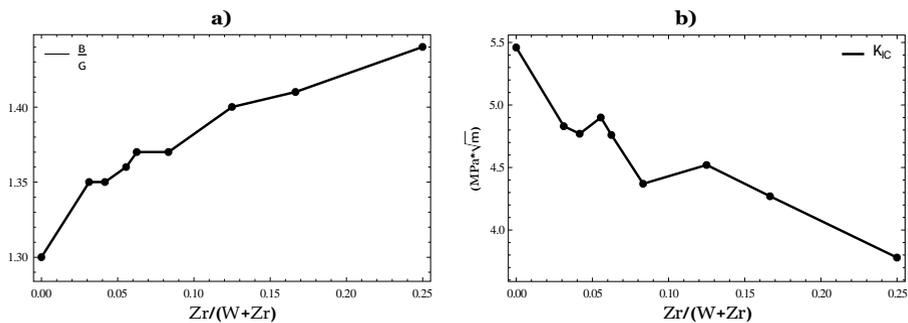

**Fig. 4** $W_{1-x}Zr_xB_2$: a) $B/G$ Pugh's ratio, b) fracture toughness $K_{IC}$ (MPa$\sqrt{m}$)

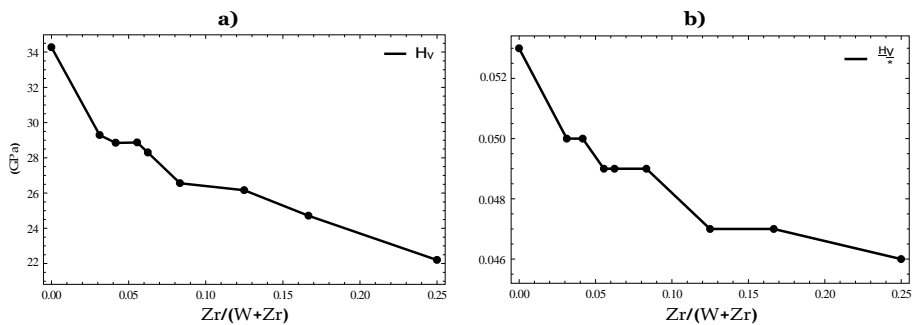

**Fig. 5** $W_{1-x}Zr_xB_2$: a) hardness $H_v$ (GPa), b) hardness to modified Young's modulus ratio $H_v/E*$

## 3.2 Comparison with experiment

In Fig.6 the exemplary results of SEM investigations are shown. The surface of deposited films is smooth and cross-section is uniform of 2.8 $\mu$m in thickness (Fig.6a,b).

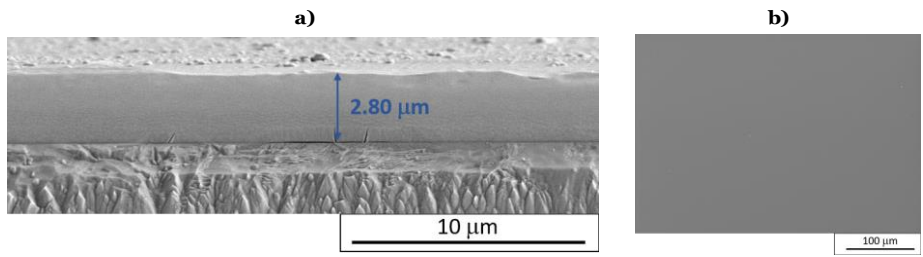

**Fig. 6** SEM investigations of coating ($W_{0.84}Zr_{0.16}B_{1.52}$) deposited from target with 24% at. of zirconium on nitrided QRO90 steel substrates: a) cross-section, b) surface

In the Tab.4 chemical composition of deposited films is presented. Deposited films are characterized by similar stoichiometric composition as predicted theoretically structures $W_{0.835}Zr_{0.165}B_2$, $W_{0.875}Zr_{0.125}B_2$ and $W_{0.915}Zr_{0.085}B_2$ respectively for 24% at., 16% at., 8% at. of $Zr$ in target. However, a marked decrease in the amount of boron in films is noteworthy. Such phenomena was observed in earlier studies [54] and can be explained be scattering of light boron atoms on heavy tungsten in plasma plume and next the resputtering of deposited boron by heavier atoms from coating. Also oxygen is detected in coatings and its content grows with an increase of zirconium amount.

**Table 4** Chemical composition of deposited coatings

| $\frac{Zr}{W+Zr}$ in target | $B$(% at.) | $Zr$(% at.) | $W$(% at.) | $\frac{Zr}{W+Zr}$ | $\frac{B}{W+Zr}$ | $O$(% at.) |
|---|---|---|---|---|---|---|
| 0.00 | 59.43 | 0.00 | 37.32 | 0.00 | 1.59 | 3.25 |
| 0.08 | 55.23 | 2.89 | 38.04 | 0.07 | 1.35 | 3.84 |
| 0.16 | 56.72 | 4.63 | 34.73 | 0.12 | 1.44 | 3.92 |
| 0.24 | 57.98 | 5.93 | 32.10 | 0.16 | 1.52 | 3.99 |

The theoretically obtained mechanical properties are lower than these from micro-hardness tests (Fig.7a). Measured with load 10 g hardness of sample $W_{0.84}Zr_{0.16}B_{1.52}$ deposited from target with 24% at. of zirconium ($W_{0.84}Zr_{0.16}B_{1.52}$) is 29.9 $\pm$ 2.4 GPa when calculated $W_{0.835}Zr_{0.165}B_2$ is characterized by $H_v$=24.71 GPa. Taking into account measurement errors (errors bars in Fig.7a) for zirconium alloyed samples the value of hardness does not change substantially and average hardness is ~32 GPa. The lowest value was obtained for undoped coating (26.8 GPa) when in theoretical calculations $WB_2$ possesses hardness 34.27 GPa. It can be explained by the fact that tungsten borides crystallize in various phases and $WB_2$ hP6-P6$_3$/mmc (194) which seems to be the hardest one [32] was chosen for calculations. Later studies have shown that the dominant phase is the deposited coatings is much softer phase P6/mmm. Addition of zirconium causes rebuilding of crystal structure and in consequence change of mechanical properties. The increase of zirconium amount results in the grow of fracture toughness $K_{IC}$ (Fig.7b) and for $W_{0.84}Zr_{0.16}B_{1.52}$ differences are the lowest and $K_{IC}$ is $3.86\pm0.15$ MPa$\sqrt{m}$ when in theory it is 4.27 MPa$\sqrt{m}$.

The differences in mechanical properties can be explained both on atomic level and based on microstructure also. The mechanical properties of the ternaries are highly sensitive to the vacancy concentration [52]. It is clearly seen that during deposition the high amount of boron is losing. It caused that crystal lattice is built with vacancies. Vacancies result in a reduction of Young's modulus, can decrease lattice parameter $c$ and may even result in a strengthening of $\alpha$-$WB_{2-z}$ [55]. It should be noted that in above presented calculations the vacancies were not taken into account. The dominating strengthening mechanism can be related to solid solution hardening effects. This mechanism consists of parelastic and dielastic contributions [11] which is resulting from the different lattice parameters and shear moduli of $ZrB_2$ and $WB_2$.

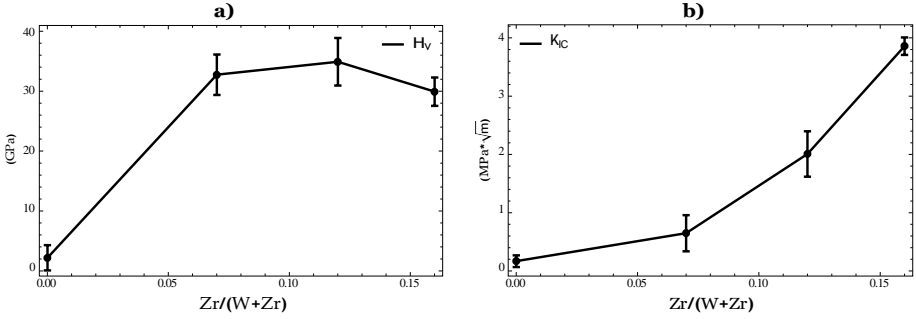

**Fig. 7** Deposited $W-Zr-B_x$ films a) hardness $H_v$ (GPa), b) fracture toughness $K_{IC}$ (MPa$\sqrt{m}$)

In Fig.8 the XRD spectrum of the coatings deposited on Si (001) revealed main diffraction peaks at ranging from $2\Theta=20°$ to $38°$. In the case of $WB_{2-z}$ coating the peak positioned at $28.9°$ comes from the (0001) plane of hexagonal $AlB_2$-type $WB_2$ ($\alpha$-$WB_2$) and the peak positioned at $26°$ is derived from the (0004) plane of hexagonal $MoB_2$-type $WB_2$ ($\omega$-$W_2B_5$). Based on detailed deconvolution of XRD spectra the $\alpha$-$WB_2$ to $\omega$-$W_2B_5$ ratio is 4.8 and both phases have a similar crystallite size of $37\pm2$ nm (calculated on the basis of Scherrer formula). In the case of $Zr$-doped coatings the $ZrB_2$ phase doesn't appear, and the diffraction lines (related to $\alpha$-$WB_2$ and $\omega$-$W_2B_5$) are shifted towards smaller $2\Theta$ angle due to higher radius of zirconium than tungsten. The shift toward smaller angles increase with $Zr$ content. Moreover, the $\alpha$-$WB_2$ to $\omega$-$W_2B_5$ ratio decrease with $Zr$. Similar behavior of alloyed $WB_{2-z}$ coating has been already observed by Moraes et al. [3]. Researchers indicate that the shift in diffraction lines is related to boron vacancies and the formation of a new phase, in our case the $W\_Zr\_B$ phase. The crystallite size of $\alpha$-$WB_2$ and $\omega$-$W_2B_5$ phases does not change significantly with $Zr$ content and is 30 and 40 nm, respectively. In the XRD spectrum, apart from narrow diffraction lines, a broad diffraction line at $2\Theta$ between $22°$ and $46°$ was observed, indicating amorphous phase.

The presence of amorphous phase is also confirmed by the STEM images that were taken in selective area electron diffraction (SAED) mode (Fig.9). There are not visible diffraction spots and mainly the amorphous halo is observed. Similar results were obtained by Moscicki et al. [54] for WB2 films alloyed with titanium. The presence of the amorphous structure in $W\_Ti\_B_{4-x}$ was confirmed by the additional fast Fourier transform (FTT) where blurred diffraction rings was recorded [54].

The higher H values of deposited coatings are primarily attributed to solid-solution hardening and their narrow columns (*Hall–Petch* effect) [58]. In the

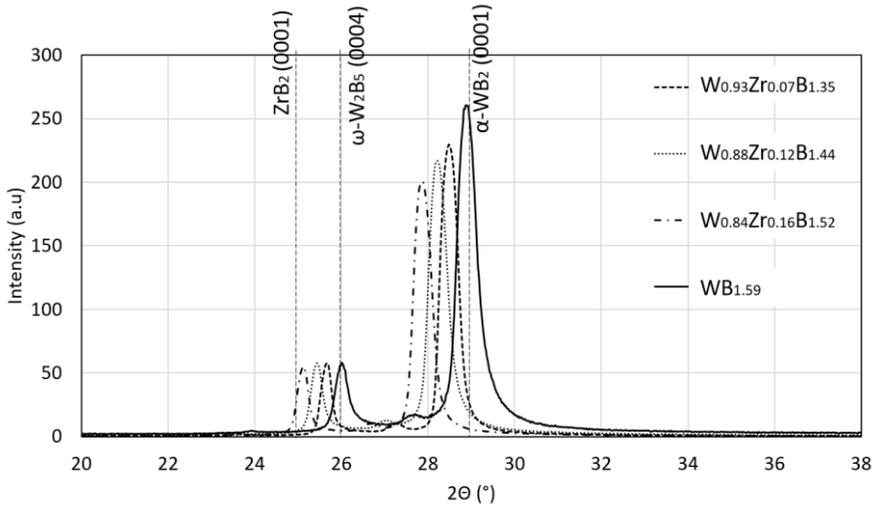

**Fig. 8** Structural evolution of the $W_{1-x}Zr_xB_{2-z}$ coatings with increasing $Zr$ content ($x$ = 0.00, 0.07, 0.12, 0.16). The standardized 2Θ-peak positions of $\omega$-$W_2B_5$ ($a$=2.983Å, $c$=13.879Å) [56], $\alpha$-$WB_2$ ($a$=3.020, $c$=3.050), $\alpha$-$ZrB_2$ ($a$=3.170Å, $c$=3.548Å) [57] are indicated with a dashed line.

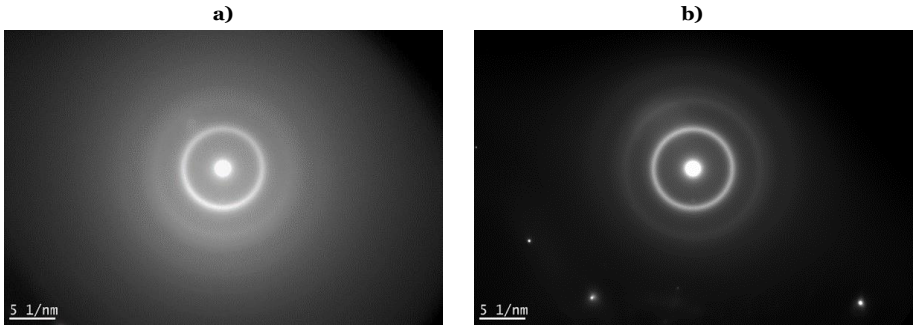

**Fig. 9** STEM images in selective area electron diffraction (SAED) mode a) $WB_{1.59}$ and b) $W_{0.93}Zr_{0.07}B_{1.35}$

figure 10 STEM images of $WB_{1.59}$ and $W_{0.93}Zr_{0.07}B_{1.35}$ coating's cross-section is shown. The obtained structure confirms earlier calculations of crystallite size. The deposited coating is characterized by a columnar structure. For $WB_{1.59}$ column are narrow (~40 nm) and perpendicular to substrate surface (Fig.10a).

The columnar structure causes material strength increase. The grain boundary strengthening effect plays a major role here. The reinforcing role of grain boundaries is that they act as barriers to dislocation movement, causing them to pile up. Plastic deformation cannot continue when the stresses reach the value necessary to initiate slip in the adjacent grain. The addition of a small amount of zirconium caused that direction of columns are changed and grains with letter "V" shape (Fig.10b) are formed. Due to structural zone models (SZM) of Barna and Adamik [59] such a shape is characteristic for "zone T" created when content of the impurities or additives is growing. The introducing of irregular direction of columns increase strength also in other directions, what can cause the growth of cracking resistance ($K_{IC}$).

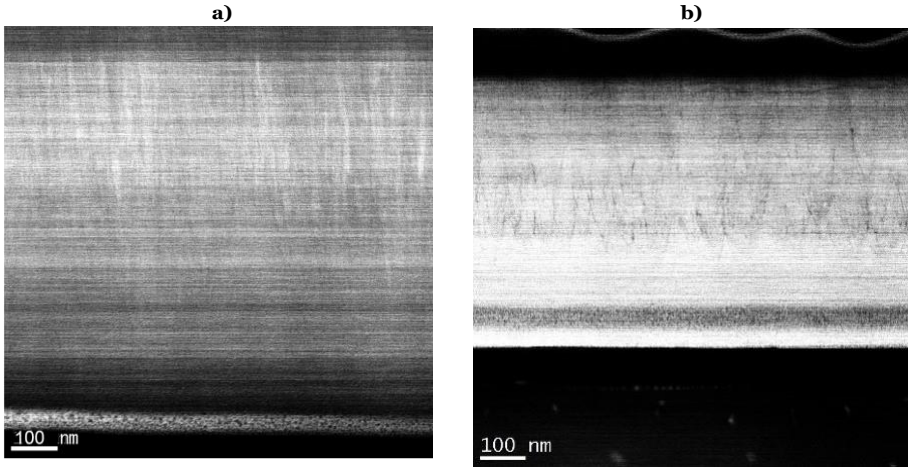

**Fig. 10** BF STEM images of cross-section of deposited layers a) $WB_{1.59}$ and b) $W_{0.93}Zr_{0.07}B_{1.35}$

## 4 Summary

Using quantum-mechanical calculations, the effect of *Zr* doping on the mechanical properties of $W_{1-x}Zr_xB_2$ was estimated. Obtained results of calculations were compared with experimental values measured on RF-magnetron sputtered coatings with similar composition. Deposited with zirconium admixture films are characterized by greater hardness $H_v$ but lower fracture toughness $K_{IC}$ than theoretical structures. Taking into account that calculation were made on ideal cell and there are several hardening mechanisms (not only solid solution hardening) the differences between calculations and experiment can be great. Additionally, not included in calculations vacancies on boron position and also change of microstructure from columnar perpendicular to substrate to grains with letter "V" shape influences the properties. However, obtained results predict new material that due to its high hardness and improved brittle-ductile character can be competitive to nowadays used materials.

It can be concluded that:

- *Zr* doped tungsten diboride is mechanically and thermodynamically stable;
- Zirconium doping reduces most mechanical properties of $W_{1-x}Zr_xB_2$, but can influence on structure of deposited films;
- modified tungsten diboride can be more ductile.

We hope that the results presented here will also be confirmed by other authors. Knowledge of zirconium doping might be useful in the design of new materials.

**Acknowledgments.** This work was supported by the National Science Centre (NCN – Poland) Research Project: UMO-2017/25/B/ST8/01789 and project SUPERCOAT; project number: TECHMASTRATEG-III/0017/2019. Additional assistance was granted through the computing cluster GRAFEN at Biocentrum Ochota, the Interdisciplinary Centre for Mathematical and Computational Modelling of Warsaw University (ICM UW) and Poznań Supercomputing and Networking Center (PSNC).

# Crystallographic information and stiffness tensors of $W_{1-x}Zr_xB_2$

## $WB_2$

```
#WB2

_symmetry_space_group_name_H-M "P 63/m 2/m 2/c"
_symmetry_Int_Tables_number 194

_cell_length_a      2.89773
_cell_length_b      2.89773
_cell_length_c      7.67174
_cell_angle_alpha   90.00000
_cell_angle_beta    90.00000
_cell_angle_gamma   120.00000
_cell_volume        55.786499

loop_
_space_group_symop_id
_space_group_symop_operation_xyz
1  x,y,z
2  x-y,x,z+1/2
3  -y,x-y,z
4  -x,-y,z+1/2
5  -x+y,-x,z
6  y,-x+y,z+1/2
7  x-y,-y,-z
8  x,x-y,-z+1/2
9  y,x,-z
10 -x+y,y,-z+1/2
11 -x,-x+y,-z
12 -y,-x,-z+1/2
13 -x,-y,-z
```

```
14  -x+y,-x,-z+1/2
15  y,-x+y,-z
16  x,y,-z+1/2
17  x-y,x,-z
18  -y,x-y,-z+1/2
19  -x+y,y,z
20  -x,-x+y,z+1/2
21  -y,-x,z
22  x-y,-y,z+1/2
23  x,x-y,z
24  y,x,z+1/2

loop_
_atom_site_label
_atom_site_type_symbol
_atom_site_symmetry_multiplicity
_atom_site_Wyckoff_label
_atom_site_fract_x
_atom_site_fract_y
_atom_site_fract_z
_atom_site_occupancy
B1  B   4  f  0.33333  0.66667  0.04149  1.00000
W1  W   2  d  0.33333  0.66667  0.75000  1.00000
```

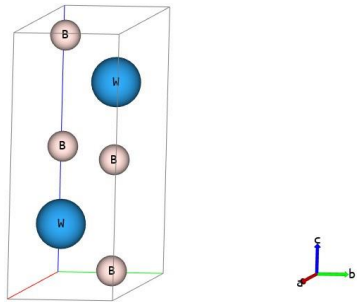

**Fig. 1** $WB_2$: Basic cell

**Stiffness tensor:**

$$[C_{IJ}] \rightarrow \begin{bmatrix} 610.214 & 206.739 & 117.767 & 0. & 0. & 0. \\ 206.739 & 610.214 & 117.767 & 0. & 0. & 0. \\ 117.767 & 117.767 & 953.482 & 0. & 0. & 0. \\ 0. & 0. & 0. & 276.306 & 0. & 0. \\ 0. & 0. & 0. & 0. & 276.306 & 0. \\ 0. & 0. & 0. & 0. & 0. & 201.737 \end{bmatrix} [\text{GPa}].$$

## $W_{0.75}Zr_{0.25}B_2$

# W3Zr1B8

_symmetry_space_group_name_H-M "P m m 2"
_symmetry_Int_Tables_number  25

_cell_length_a      2.92850
_cell_length_b      7.86539
_cell_length_c      5.09269
_cell_angle_alpha   90.00000
_cell_angle_beta    90.00000
_cell_angle_gamma   90.00000
_cell_volume        117.303970

loop_
_space_group_symop_id
_space_group_symop_operation_xyz
1  x,y,z
2  -x,-y,z
3  -x,y,z
4  x,-y,z

loop_
_atom_site_label
_atom_site_type_symbol
_atom_site_symmetry_multiplicity
_atom_site_Wyckoff_label
_atom_site_fract_x
_atom_site_fract_y
_atom_site_fract_z
_atom_site_occupancy
B1   B    2 h  0.50000  0.20119  0.16510  1.00000
B2   B    2 g  0.00000  0.72339  0.34035  1.00000
Zr1  Zr   1 d  0.50000  0.50000  0.16899  1.00000
W1   W    1 a  0.00000  0.00000  0.33328  1.00000
B3   B    2 g  0.00000  0.20557  0.66808  1.00000
B4   B    2 h  0.50000  0.71599  0.82637  1.00000
W2   W    1 b  0.00000  0.50000  0.66423  1.00000
W3   W    1 c  0.50000  0.00000  0.83369  1.00000

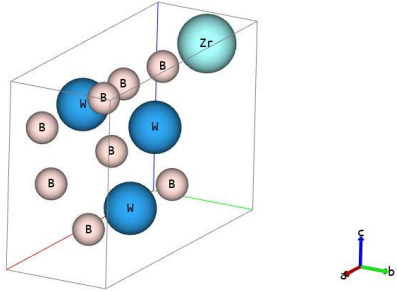

**Fig. 2** $W_{0.75}Zr_{0.25}B_2$: Basic cell

**Stiffness tensor:**

$$[C_{IJ}] \rightarrow \begin{bmatrix} 478.007 & 182.029 & 104.013 & 0. & 0. & 0. \\ 182.029 & 460.394 & 100.821 & 0. & 0. & 0. \\ 104.013 & 100.821 & 778.679 & 0. & 0. & 0. \\ 0. & 0. & 0. & 199.078 & 0. & 0. \\ 0. & 0. & 0. & 0. & 200.996 & 0. \\ 0. & 0. & 0. & 0. & 0. & 143.843 \end{bmatrix} [GPa].$$

## .1 $W_{0.835}Zr_{0.165}B_2$

# W5Zr1B12

```
_symmetry_space_group_name_H-M "P 1 m 1"
_symmetry_Int_Tables_number  6

_cell_length_a     2.92020
_cell_length_b     7.79490
_cell_length_c     7.74161
_cell_angle_alpha  90.00000
_cell_angle_beta   100.86151
_cell_angle_gamma  90.00000
_cell_volume       173.062812

loop_
_space_group_symop_id
_space_group_symop_operation_xyz
1 x,y,z
2 x,-y,z

loop_
_atom_site_label
_atom_site_type_symbol
```

```
_atom_site_symmetry_multiplicity
_atom_site_Wyckoff_label
_atom_site_fract_x
_atom_site_fract_y
_atom_site_fract_z
_atom_site_occupancy
B1   B    2 c 0.55506 0.70158 0.11007 1.00000
B2   B    2 c 0.11408 0.22189 0.22820 1.00000
Zr1  Zr   1 a 0.55643 0.00000 0.11321 1.00000
W1   W    1 b 0.11201 0.50000 0.22382 1.00000
B3   B    2 c 0.22357 0.70665 0.44711 1.00000
B4   B    2 c 0.77735 0.21081 0.55471 1.00000
W2   W    1 a 0.22479 0.00000 0.44974 1.00000
W3   W    1 b 0.77804 0.50000 0.55587 1.00000
B5   B    2 c 0.88801 0.70716 0.77598 1.00000
B6   B    2 c 0.44182 0.21480 0.88365 1.00000
W4   W    1 a 0.88523 0.00000 0.77078 1.00000
W5   W    1 b 0.44373 0.50000 0.88715 1.00000
```

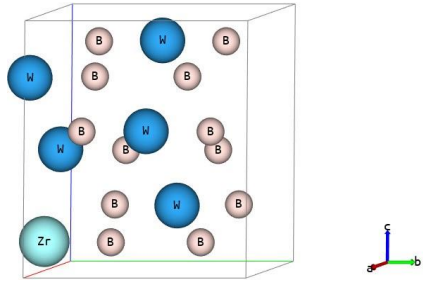

**Fig. 3** $W_{0.835}Zr_{0.165}B_2$: Basic cell

**Stiffness tensor:**

$$[C_{IJ}] \rightarrow \begin{bmatrix} 492.413 & 193.593 & 107.138 & 0. & 0. & 9.467 \\ 193.593 & 523.859 & 113.458 & 0. & 0. & 0.231 \\ 107.138 & 113.458 & 846.563 & 0. & 0. & -0.362 \\ 0. & 0. & 0. & 224.912 & -0.623 & 0. \\ 0. & 0. & 0. & -0.623 & 217.7 & 0. \\ 9.467 & 0.231 & -0.362 & 0. & 0. & 157.822 \end{bmatrix} \text{[GPa]}.$$

# $W_{0.875}Zr_{0.125}B_2$

# W7Zr1B16

```
_symmetry_space_group_name_H-M "A m m 2"
_symmetry_Int_Tables_number 38

_cell_length_a      7.76000
_cell_length_b      5.83172
_cell_length_c      10.11121
_cell_angle_alpha   90.00000
_cell_angle_beta    90.00000
_cell_angle_gamma   90.00000
_cell_volume        457.574184

loop_
_space_group_symop_id
_space_group_symop_operation_xyz
1  x,y,z
2  -x,-y,z
3  x,-y,z
4  -x,y,z
5  x,y+1/2,z+1/2
6  -x,-y+1/2,z+1/2
7  x,-y+1/2,z+1/2
8  -x,y+1/2,z+1/2

loop_
_atom_site_label
_atom_site_type_symbol
_atom_site_symmetry_multiplicity
_atom_site_Wyckoff_label
_atom_site_fract_x
_atom_site_fract_y
_atom_site_fract_z
_atom_site_occupancy
B1   B    8 f  0.29282  0.25106   0.41635  1.00000
B2   B    8 f  0.78400  0.74474   0.08156  1.00000
W1   W    4 d  0.00000  0.74875  -0.08290  1.00000
W2   W    4 e  0.50000  0.25016   0.58337  1.00000
B3   B    4 c  0.29283  0.00000   0.66740  1.00000
B4   B    4 c  0.79218  0.00000   0.83330  1.00000
W3   W    2 a  0.00000  0.00000   0.66585  1.00000
W4   W    2 b  0.50000  0.00000   0.83332  1.00000
B5   B    4 c  0.29747  0.00000   0.16668  1.00000
B6   B    4 c  0.78398  0.00000   0.33681  1.00000
Zr1  Zr   2 a  0.00000  0.00000   0.16669  1.00000
W5   W    2 b  0.50000  0.00000   0.33320  1.00000
```

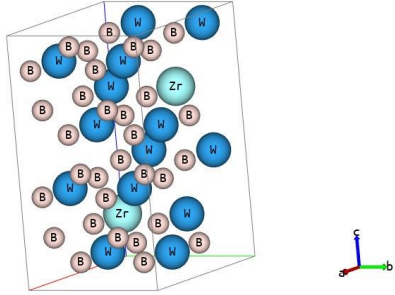

**Fig. 4**  $W_{0.875}Zr_{0.125}B_2$: Basic cell

**Stiffness tensor:**

$$[C_{IJ}] \to \begin{bmatrix} 541.56 & 200.813 & 111.13 & 0. & 0. & 0. \\ 200.813 & 532.753 & 112.813 & 0. & 0. & 0. \\ 111.13 & 112.813 & 852.923 & 0. & 0. & 0. \\ 0. & 0. & 0. & 231.813 & 0. & 0. \\ 0. & 0. & 0. & 0. & 229.596 & 0. \\ 0. & 0. & 0. & 0. & 0. & 166.8 \end{bmatrix} [GPa].$$

# $W_{0.915}Zr_{0.085}B_2$

# W11Zr1B24

```
_symmetry_space_group_name_H-M "P 1 m 1"
_symmetry_Int_Tables_number  6

_cell_length_a      5.81990
_cell_length_b      7.72960
_cell_length_c      7.70561
_cell_angle_alpha   90.00000
_cell_angle_beta    100.89389
_cell_angle_gamma   90.00000
_cell_volume        340.393845

loop_
_space_group_symop_id
_space_group_symop_operation_xyz
1  x,y,z
2  x,-y,z

loop_
_atom_site_label
_atom_site_type_symbol
```

```
_atom_site_symmetry_multiplicity
_atom_site_Wyckoff_label
_atom_site_fract_x
_atom_site_fract_y
_atom_site_fract_z
_atom_site_occupancy
B1   B    2 c  0.77759   0.70726 0.11041 1.00000
B2   B    2 c  0.05107   0.21540 0.22524 1.00000
W1   W    1 a  0.77811   0.00000 0.11275 1.00000
W2   W    1 b  0.05560   0.50000 0.22304 1.00000
B3   B    2 c  0.11111   0.70770 0.44580 1.00000
B4   B    2 c  0.38864   0.20899 0.55470 1.00000
W3   W    1 a  0.10949   0.00000 0.44668 1.00000
W4   W    1 b  0.38932   0.50000 0.55473 1.00000
B5   B    2 c  0.44551   0.70785 0.77656 1.00000
B6   B    2 c  0.72216   0.20781 0.88918 1.00000
W5   W    1 a  0.44474   0.00000 0.77456 1.00000
W6   W    1 b  0.72200   0.50000 0.88888 1.00000
B7   B    2 c  0.27774   0.70280 0.11087 1.00000
B8   B    2 c  0.56161   0.21528 0.22527 1.00000
Zr1  Zr   1 a  0.27777   0.00000 0.11090 1.00000
W7   W    1 b  0.55593   0.50000 0.22290 1.00000
B9   B    2 c  0.61172   0.70767 0.44566 1.00000
B10  B    2 c  0.88896   0.21029 0.55570 1.00000
W8   W    1 a  0.61409   0.00000 0.44783 1.00000
W9   W    1 b  0.88852   0.50000 0.55665 1.00000
B11  B    2 c  -0.05709  0.70776 0.77719 1.00000
B12  B    2 c  0.22101   0.21505 0.88330 1.00000
W10  W    1 a  -0.05771  0.00000 0.77397 1.00000
W11  W    1 b  0.22210   0.50000 0.88735 1.00000
```

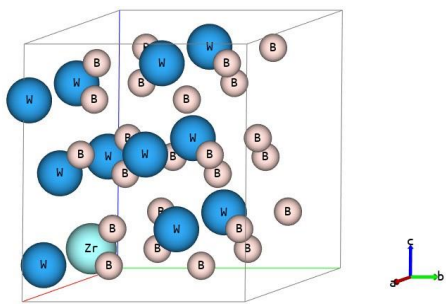

**Fig. 5** $W_{0.915}Zr_{0.085}B_2$: Basic cell

**Stiffness tensor:**

$$[C_{IJ}] \rightarrow \begin{bmatrix} 530.328 & 186.025 & 114.522 & 0. & 0. & -9.041 \\ 186.025 & 480.47 & 113.165 & 0. & 0. & -4.657 \\ 114.522 & 113.165 & 879.372 & 0. & 0. & -11.539 \\ 0. & ( . & 0. & 240.626 & -11.962 & 0. \\ 0. & ( . & 0. & -11.962 & 239.934 & 0. \\ -9.041 & -4.657 & -11.539 & 0. & 0 & 156.3 \end{bmatrix} [GPa].$$

## $W_{0.937}Zr_{0.063}B_2$

# W15Zr1B32

_symmetry_space_group_name_H-M "A m m 2"
_symmetry_Int_Tables_number 38

_cell_length_a      15.43145
_cell_length_b      5.81311
_cell_length_c      10.07829
_cell_angle_alpha   90.00000
_cell_angle_beta    90.00000
_cell_angle_gamma   90.00000
_cell_volume        904.0701

loop_
_space_group_symop_id
_space_group_symop_operation_xyz
1  x,y,z
2  -x,-y,z
3  x,-y,z
4  -x,y,z
5  x,y+1/2,z+1/2
6  -x,-y+1/2,z+1/2
7  x,-y+1/2,z+1/2
8  -x,y+1/2,z+1/2

loop_
_atom_site_label
_atom_site_type_symbol
_atom_site_symmetry_multiplicity
_atom_site_Wyckoff_label
_atom_site_fract_x
_atom_site_fract_y
_atom_site_fract_z
_atom_site_occupancy

| | | | | | | |
|---|---|---|---|---|---|---|
| B1 | B | 8 f | 0.14734 | 0.25081 | 0.41643 | 1.00000 |
| B2 | B | 8 f | 0.89112 | 0.74498 | 0.08163 | 1.00000 |
| B3 | B | 8 f | 0.35518 | 0.25026 | 0.41660 | 1.00000 |
| W1 | W | 4 d | 0.00000 | 0.74863 | -0.08287 | 1.00000 |
| W2 | W | 8 f | 0.25182 | 0.75006 | 0.08334 | 1.00000 |
| B4 | B | 4 c | 0.14734 | 0.00000 | 0.66722 | 1.00000 |
| B5 | B | 4 c | 0.89566 | 0.00000 | 0.83330 | 1.00000 |
| B6 | B | 4 c | 0.35518 | 0.00000 | 0.66687 | 1.00000 |
| W3 | W | 2 a | 0.00000 | 0.00000 | 0.66576 | 1.00000 |
| W4 | W | 4 c | 0.25085 | 0.00000 | 0.83332 | 1.00000 |
| B7 | B | 4 c | 0.15001 | 0.00000 | 0.16668 | 1.00000 |
| B8 | B | 4 c | 0.89112 | 0.00000 | 0.33665 | 1.00000 |
| B9 | B | 4 c | 0.35554 | 0.00000 | 0.16669 | 1.00000 |
| Zr1 | Zr | 2 a | 0.00000 | 0.00000 | 0.16669 | 1.00000 |
| W5 | W | 4 c | 0.25181 | 0.00000 | 0.33327 | 1.00000 |
| B10 | B | 8 f | 0.39658 | 0.74992 | 0.08329 | 1.00000 |
| W6 | W | 4 e | 0.50000 | 0.75010 | -0.08334 | 1.00000 |
| B11 | B | 4 c | 0.39620 | 0.00000 | 0.83331 | 1.00000 |
| W7 | W | 2 b | 0.50000 | 0.00000 | 0.66675 | 1.00000 |
| B12 | B | 4 c | 0.39658 | 0.00000 | 0.33336 | 1.00000 |
| W8 | W | 2 b | 0.50000 | 0.00000 | 0.16669 | 1.00000 |

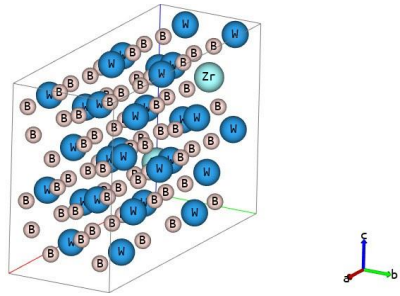

**Fig. 6** $W_{0.937}Zr_{0.063}B_2$: Basic cell

**Stiffness tensor:**

$$[C_{IJ}] \rightarrow \begin{bmatrix} 550.316 & 201.993 & 112.048 & 0. & 0. & 0. \\ 201.993 & 557.618 & 112.261 & 0. & 0. & 0. \\ 112.048 & 112.261 & 891.579 & 0. & 0. & 0. \\ 0. & 0. & 0. & 243.58 & 0. & 0. \\ 0. & 0. & 0. & 0. & 242.94 & 0. \\ 0. & 0. & 0. & 0. & 0. & 175.921 \end{bmatrix} \text{[GPa]}.$$

# $W_{0.944}Zr_{0.056}B_2$

# W17Zr1B36

_symmetry_space_group_name_H-M "A m m 2"
_symmetry_Int_Tables_number 38

_cell_length_a      7.71029
_cell_length_b      8.71823
_cell_length_c      15.10683
_cell_angle_alpha   90.00000
_cell_angle_beta    90.00000
_cell_angle_gamma   90.00000
_cell_volume        1015.4823

loop_
_space_group_symop_id
_space_group_symop_operation_xyz
1 x,y,z
2 -x,-y,z
3 x,-y,z
4 -x,y,z
5 x,y+1/2,z+1/2
6 -x,-y+1/2,z+1/2
7 x,-y+1/2,z+1/2
8 -x,y+1/2,z+1/2

loop_
_atom_site_label
_atom_site_type_symbol
_atom_site_symmetry_multiplicity
_atom_site_Wyckoff_label
_atom_site_fract_x
_atom_site_fract_y
_atom_site_fract_z
_atom_site_occupancy
B1  B   8 f 0.29218 0.33311  0.44380 1.00000
B2  B   8 f 0.78523 0.82923  0.05418 1.00000
W1  W   4 d 0.00000 0.83127 -0.05695 1.00000
W2  W   4 e 0.50000 0.33237  0.55523 1.00000
B3  B   8 f 0.29220 0.66557  0.11133 1.00000
B4  B   8 f 0.79170 0.66674  0.22219 1.00000
W3  W   4 d 0.00000 0.16352  0.61079 1.00000
W4  W   4 e 0.50000 0.16644  0.72228 1.00000
B5  B   4 c 0.29151 0.00000  0.77778 1.00000
B6  B   4 c 0.79170 0.00000  0.88893 1.00000

```
W5   W    2 a  0.00000  0.00000  0.77780  1.00000
W6   W    2 b  0.50000  0.00000  0.88874  1.00000
B7   B    8 f  0.29219  0.16754  0.27822  1.00000
B8   B    8 f  0.79023  0.16613  0.38906  1.00000
W7   W    4 d  0.00000  0.66775  0.77953  1.00000
W8   W    4 e  0.50000  0.66662  0.88890  1.00000
B9   B    4 c  0.29180  0.00000  0.44445  1.00000
B10  B    4 c  0.79024  0.00000  0.55519  1.00000
W9   W    2 a  0.00000  0.00000  0.44445  1.00000
W10  W    2 b  0.50000  0.00000  0.55550  1.00000
B11  B    4 c  0.29703  0.00000  0.11111  1.00000
B12  B    4 c  0.78522  0.00000  0.22495  1.00000
Zr1  Zr   2 a  0.00000  0.00000  0.11113  1.00000
W11  W    2 b  0.50000  0.00000  0.22285  1.00000
```

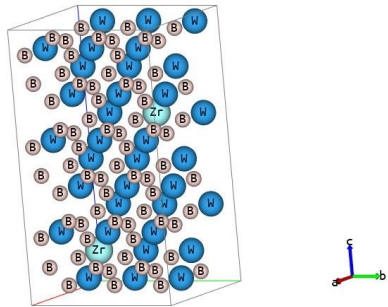

**Fig. 7** $W_{0.944}Zr_{0.056}B_2$: Basic cell

**Stiffness tensor:**

$$[C_{IJ}] \rightarrow \begin{bmatrix} 556.402 & 204.376 & 119.244 & 0. & 0. & 0. \\ 204.376 & 552.65 & 119.067 & 0. & 0. & 0. \\ 119.244 & 119.067 & 919.327 & 0. & 0. & 0. \\ 0. & 0. & 0. & 252.179 & 0. & 0. \\ 0. & 0. & 0. & 0. & 251.884 & 0. \\ 0. & 0. & 0. & 0. & 0. & 179.463 \end{bmatrix} [GPa].$$

## $W_{0.958}Zr_{0.042}B_2$

# W23Zr1B48

_symmetry_space_group_name_H-M "P 1 m 1"
_symmetry_Int_Tables_number  6

_cell_length_a       5.85400

```
_cell_length_b      15.49600
_cell_length_c      7.74411
_cell_angle_alpha   90.00000
_cell_angle_beta    100.89339
_cell_angle_gamma   90.00000
_cell_volume        689.837402

loop_
_space_group_symop_id
_space_group_symop_operation_xyz
1 x,y,z
2 x,-y,z

loop_
_atom_site_label
_atom_site_type_symbol
_atom_site_symmetry_multiplicity
_atom_site_Wyckoff_label
_atom_site_fract_x
_atom_site_fract_y
_atom_site_fract_z
_atom_site_occupancy
B1   B    2 c  0.77778   0.35450  0.11111  1.00000
B2   B    2 c  0.05555   0.60450  0.22222  1.00000
B3   B    2 c  0.77778   0.14550  0.11111  1.00000
W1   W    1 b  0.77778   0.50000  0.11111  1.00000
W2   W    2 c  0.05555   0.25000  0.22222  1.00000
B4   B    2 c  0.11111   0.35450  0.44444  1.00000
B5   B    2 c  0.38889   0.60450  0.55556  1.00000
B6   B    2 c  0.11111   0.14550  0.44444  1.00000
W3   W    1 b  0.11111   0.50000  0.44444  1.00000
W4   W    2 c  0.38889   0.25000  0.55556  1.00000
B7   B    2 c  0.44445   0.35450  0.77778  1.00000
B8   B    2 c  0.72222   0.60450  0.88889  1.00000
B9   B    2 c  0.44445   0.14550  0.77778  1.00000
W5   W    1 b  0.44445   0.50000  0.77778  1.00000
W6   W    2 c  0.72222   0.25000  0.88889  1.00000
B10  B    2 c  0.27778   0.35450  0.11111  1.00000
B11  B    2 c  0.55555   0.60450  0.22222  1.00000
B12  B    2 c  0.27778   0.14550  0.11111  1.00000
Zr1  Zr   1 b  0.27778   0.50000  0.11111  1.00000
W7   W    2 c  0.55555   0.25000  0.22222  1.00000
B13  B    2 c  0.61111   0.35450  0.44444  1.00000
B14  B    2 c  0.88889   0.60450  0.55556  1.00000
B15  B    2 c  0.61111   0.14550  0.44444  1.00000
W8   W    1 b  0.61111   0.50000  0.44444  1.00000
```

| | | | | | | |
|---|---|---|---|---|---|---|
| W9 | W | 2 | c | 0.88889 | 0.25000 | 0.55556 1.00000 |
| B16 | B | 2 | c | -0.05555 | 0.35450 | 0.77778 1.00000 |
| B17 | B | 2 | c | 0.22222 | 0.60450 | 0.88889 1.00000 |
| B18 | B | 2 | c | -0.05555 | 0.14550 | 0.77778 1.00000 |
| W10 | W | 1 | b | -0.05555 | 0.50000 | 0.77778 1.00000 |
| W11 | W | 2 | c | 0.22222 | 0.25000 | 0.88889 1.00000 |
| B19 | B | 2 | c | 0.05555 | 0.10450 | 0.22222 1.00000 |
| W12 | W | 1 | a | 0.77778 | 0.00000 | 0.11111 1.00000 |
| B20 | B | 2 | c | 0.38889 | 0.10450 | 0.55556 1.00000 |
| W13 | W | 1 | a | 0.11111 | 0.00000 | 0.44444 1.00000 |
| B21 | B | 2 | c | 0.72222 | 0.10450 | 0.88889 1.00000 |
| W14 | W | 1 | a | 0.44445 | 0.00000 | 0.77778 1.00000 |
| B22 | B | 2 | c | 0.55555 | 0.10450 | 0.22222 1.00000 |
| W15 | W | 1 | a | 0.27778 | 0.00000 | 0.11111 1.00000 |
| B23 | B | 2 | c | 0.88889 | 0.10450 | 0.55556 1.00000 |
| W16 | W | 1 | a | 0.61111 | 0.00000 | 0.44444 1.00000 |
| B24 | B | 2 | c | 0.22222 | 0.10450 | 0.88889 1.00000 |
| W17 | W | 1 | a | -0.05555 | 0.00000 | 0.77778 1.00000 |

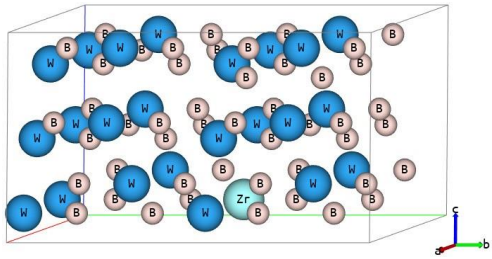

**Fig. 8** $W_{0.958}Zr_{0.042}B_2$: Basic cell

**Stiffness tensor:**

$$[C_{IJ}] \rightarrow \begin{bmatrix} 537.192 & 197.719 & 117.261 & 0. & 0. & -6.457 \\ 197.719 & 545.365 & 112.046 & 0. & 0. & -12.885 \\ 117.261 & 112.046 & 910.756 & 0. & 0. & -0.329 \\ 0. & 0. & 0. & 248.433 & -0.956 & 0. \\ 0. & 0. & 0. & -0.956 & 252.497 & 0. \\ -6.457 & -12.885 & -0.329 & 0. & 0. & 174.342 \end{bmatrix} [GPa].$$

## $W_{0.969}Zr_{0.031}B_2$

# W31Zr1B64

```
_symmetry_space_group_name_H-M "A m m 2"
_symmetry_Int_Tables_number  38

_cell_length_a      7.69207
_cell_length_b      11.61541
_cell_length_c      20.11349
_cell_angle_alpha 90.00000
_cell_angle_beta   90.00000
_cell_angle_gamma 90.00000
_cell_volume       1797.077344

loop_
_space_group_symop_id
_space_group_symop_operation_xyz
1  x,y,z
2  -x,-y,z
3  x,-y,z
4  -x,y,z
5  x,y+1/2,z+1/2
6  -x,-y+1/2,z+1/2
7  x,-y+1/2,z+1/2
8  -x,y+1/2,z+1/2

loop_
_atom_site_label
_atom_site_type_symbol
_atom_site_symmetry_multiplicity
_atom_site_Wyckoff_label
_atom_site_fract_x
_atom_site_fract_y
_atom_site_fract_z
_atom_site_occupancy
B1  B   8 f 0.29214 0.37456  0.45764  1.00000
B2  B   8 f 0.78531 0.87170  0.04057  1.00000
W1  W   4 d 0.00000 0.87339 -0.04271  1.00000
W2  W   4 e 0.50000 0.37399  0.54134  1.00000
B3  B   8 f 0.29214 0.74874  0.08346  1.00000
B4  B   8 f 0.79190 0.75000  0.16667  1.00000
W3  W   4 d 0.00000 0.24762  0.58305  1.00000
W4  W   4 e 0.50000 0.24975  0.66675  1.00000
B5  B   8 f 0.29145 0.62487  0.20837  1.00000
B6  B   8 f 0.79132 0.62495  0.29165  1.00000
W5  W   4 d 0.00000 0.12476  0.70841  1.00000
W6  W   4 e 0.50000 0.12487  0.79163  1.00000
B7  B   4 c 0.29145 0.00000  0.83325  1.00000
B8  B   4 c 0.79190 0.00000 -0.08333  1.00000
```

| | | | | | | | |
|---|---|---|---|---|---|---|---|
| W7  | W  | 2 | a | 0.00000 | 0.00000 | 0.83317  | 1.00000 |
| W8  | W  | 2 | b | 0.50000 | 0.00000 | -0.08349 | 1.00000 |
| B9  | B  | 8 | f | 0.29157 | 0.25012 | 0.33329  | 1.00000 |
| B10 | B  | 8 | f | 0.79094 | 0.24937 | 0.41648  | 1.00000 |
| W9  | W  | 4 | d | 0.00000 | 0.75000 | 0.83333  | 1.00000 |
| W10 | W  | 4 | e | 0.50000 | 0.74939 | -0.08374 | 1.00000 |
| B11 | B  | 8 | f | 0.29157 | 0.12473 | 0.45824  | 1.00000 |
| B12 | B  | 8 | f | 0.79094 | 0.62439 | 0.04145  | 1.00000 |
| W11 | W  | 4 | d | 0.00000 | 0.62450 | -0.04183 | 1.00000 |
| W12 | W  | 4 | e | 0.50000 | 0.12408 | 0.54157  | 1.00000 |
| B13 | B  | 4 | c | 0.29157 | 0.00000 | 0.58341  | 1.00000 |
| B14 | B  | 4 | c | 0.79132 | 0.00000 | 0.66671  | 1.00000 |
| W13 | W  | 2 | a | 0.00000 | 0.00000 | 0.58333  | 1.00000 |
| W14 | W  | 2 | b | 0.50000 | 0.00000 | 0.66676  | 1.00000 |
| B15 | B  | 8 | f | 0.29214 | 0.12582 | 0.20890  | 1.00000 |
| B16 | B  | 8 | f | 0.79094 | 0.12498 | 0.29208  | 1.00000 |
| W15 | W  | 4 | d | 0.00000 | 0.62577 | 0.70965  | 1.00000 |
| W16 | W  | 4 | e | 0.50000 | 0.62532 | 0.79218  | 1.00000 |
| B17 | B  | 4 | c | 0.29157 | 0.00000 | 0.33351  | 1.00000 |
| B18 | B  | 4 | c | 0.79114 | 0.00000 | 0.41667  | 1.00000 |
| W17 | W  | 2 | a | 0.00000 | 0.00000 | 0.33366  | 1.00000 |
| W18 | W  | 2 | b | 0.50000 | 0.00000 | 0.41667  | 1.00000 |
| B19 | B  | 4 | c | 0.29718 | 0.00000 | 0.08333  | 1.00000 |
| B20 | B  | 4 | c | 0.78532 | 0.00000 | 0.16887  | 1.00000 |
| Zr1 | Zr | 2 | a | 0.00000 | 0.00000 | 0.08333  | 1.00000 |
| W19 | W  | 2 | b | 0.50000 | 0.00000 | 0.16735  | 1.00000 |

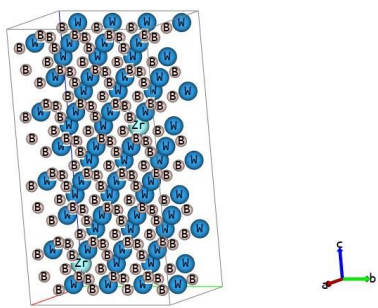

**Fig. 9** $W_{0.969}Zr_{0.031}B_2$: Basic cell

**Stiffness tensor:**

$$[\mathbf{C_{IJ}}] \rightarrow \begin{bmatrix} 538.607 & 197.882 & 116.845 & 0. & 0. & 0. \\ 197.882 & 551.257 & 117.099 & 0. & 0. & 0. \\ 116.845 & 117.099 & 926.041 & 0. & 0. & 0. \\ 0. & 0. & 0. & 254.858 & 0. & 0. \\ 0. & 0. & 0. & 0. & 254.479 & 0. \\ 0. & 0. & 0. & 0. & 0. & 175.458 \end{bmatrix} [\text{GPa}].$$